\documentclass{optica-article}

\journal{opticajournal} 

\articletype{Research Article}

\usepackage{lineno}

\begin{document}

\title{Bright dual-pulse betatron X-ray generation from a laser wakefield accelerator}

\author{Bo Guo,\authormark{1} Yang Wan,\authormark{2,3,*} Shuang Liu,\authormark{4} Xiaonan Ning,\authormark{1} Jianfei Hua,\authormark{5} and Wei Lu\authormark{4,1,5,†}}

\address{\authormark{1}Beijing Academy of Quantum Information Sciences, Beijing, 100193, China\\
\authormark{2}School of Physics and Laboratory of Zhongyuan Light, Zhengzhou University, Zhengzhou, 450001, China\\
\authormark{3}State Key Laboratory of Critical Metals Beneficiation, Metallurgy and Purification, Zhengzhou University, Zhengzhou, 450001, China\\
\authormark{4}Institute of High Energy Physics, Chinese Academy of Sciences, Beijing, 100049, China\\
\authormark{5}Department of Engineering Physics, Tsinghua University, Beijing, 100084, China}

\email{\authormark{*}yangwan23@zzu.edu.cn}
\email{\authormark{†}weilu@ihep.ac.cn}


\begin{abstract*} 
Pump-probe experiments using dual ultrashort X-ray pulses provide unique opportunities for resolving non-equilibrium dynamics initiated by intense X-ray excitation. Betatron radiation from laser wakefield accelerators offers femtosecond duration, micrometer-scale source size, and intrinsic synchronization with the driving laser, making it a promising candidate for compact ultrafast X-ray sources.
Here, we experimentally demonstrate a high-flux, dual-pulse betatron X-ray source based on a density-tailored gas-mixture target. Two electron bunches are generated within a single plasma wakefield through ionization-induced and shock-front-triggered injection, subsequently producing twin X-ray pulses. The measured electron spectra and dual-component X-ray angular profiles, together with particle-in-cell simulations, identify the contributions of the two electron populations to the radiation. The total X-ray photon yield reaches the level of $10^{10}$ photons per shot with a 40-TW laser system. These results establish a compact, single-stage route toward high-flux dual-pulse betatron sources for laboratory-scale ultrafast X-ray spectroscopy.
\end{abstract*}

\section{Introduction}
 Ultrafast pump-probe measurements based on femtosecond X-ray pulses have become a powerful tool for exploring the dynamics of atomic, molecular, and solid-state systems in modern physics, femtosecond chemistry and structural biology. In particular, the emergence of dual- and multi-pulse operation at X-ray free-electron lasers (XFELs) \cite{guo2024experimental,bostedt2016linac} has opened new possibilities beyond the conventional configuration in which an infrared laser pulse serves as a pump and only a single X-ray pulse serves as a probe. For instance, one potential type of experiment, using multiple X-ray pulses to probe non-equilibrium states at different delays after a single pump, can substantially reduce both the acquisition time and the number of destructive exposures imposed on the sample. Another potential type uses the first X-ray pulse as the pump and a subsequent X-ray pulse as the probe, enabling selective excitation of electronic resonances \cite{vinko2012creation}, thereby providing element-specific insights into ultrafast processes.
At present, this capability remains confined to a few XFEL facilities, and beamtime for nonstandard multi-pulse operation is particularly restricted. A compact dual-pulse ultrashort X-ray source that can operate in a conventional laboratory would therefore substantially broaden access to such measurements.

Betatron radiation from laser wakefield accelerators \cite{tajima1979laser,malka2008principles,faure2004laser,geddes2004high,mangles2004monoenergetic} (LWFAs) offers an attractive laboratory-scale approach. Generated by relativistic electrons undergoing transverse oscillations in a plasma wake \cite{rousse2004production,kneip2010bright,corde2013femtosecond}, betatron X-rays naturally combine femtosecond duration, micrometer-scale source size, and high peak brightness \cite{kettle2019single,dopp2017stable,cipiccia2011gamma,zhang2025high}. These properties have already enabled high-resolution phase-contrast imaging \cite{cole2018high,wenz2015quantitative,dopp2018quick,guo2019high} and ultrafast probing of warm dense matter \cite{mahieu2018probing,mo2017measurements}. 

The generation of multi-pulse betatron radiation is governed by the injection and subsequent evolution of multiple electron populations within the plasma wake. Previous experiments have generated multiple electron bunches by combining shock-front injection with colliding-pulse injection, plasma-optics-induced laser splitting, asymmetric injection geometries or wavefront-aberrated drive laser \cite{wenz2019dual,seemann2023refractive,levine2025direct,li2013observation}. In particular, Wenz et al. demonstrated tunable dual electron beams and discussed their use for downstream inverse-Compton radiation \cite{wenz2019dual}. Pulse trains of electron bunches or betatron radiation have also been investigated theoretically and numerically \cite{horny2020attosecond,jakobsson2021gev}. Separately, longitudinal and transverse density tailoring, multi-jet targets, and electron rephasing have been used to enhance the photon energy or flux of betatron sources \cite{kozlova2020hard,tomkus2020laser,rakowski2022transverse,gautier2025decoupling}. These studies established important methods for either multi-bunch generation or radiation enhancement.
However, experimentally combining dual-pulse betatron emission with a $10^{10}$-photons/shot yield remains a challenge, even though both attributes are crucial for pump–probe experiments.

Here, we report a compact, single-stage scheme that simultaneously generates high-charge dual electron bunches with enhanced betatron emission using a single 40-TW-class laser pulse. A density-tailored hydrogen-nitrogen target first supports ionization-induced injection and subsequently triggers a second electron bunch at the shock-front downramp. The two electron populations generate distinct radiation components, while rephasing acceleration inside the density structure enhances the total X-ray output. The measured photon yield reaches $10^{10}$ photons/shot, corresponding to a photon flux exceeding $10^{6}$ photons/0.1\%BW/shot at 0.5--7 keV---nearly an order of magnitude higher than that obtained in a uniform-density plasma.

\begin{figure}[htb]
\centering\includegraphics{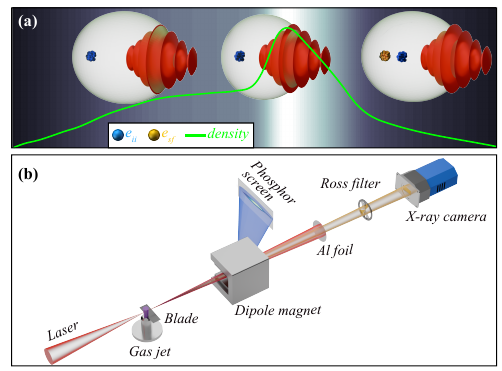}
\caption{\label{fig1} \textbf{Schematic of dual-pulse betatron X-ray generation in a density-tailored gas-mixture target.} (a) Betatron radiation from  the ionization-injected electrons, $e_{ii}$ (blue), is enhanced due to the rephasing acceleration in the density hump. At the falling edge of the shock front, a second electron bunch $e_{sf}$ (yellow) is injected by the shock-front-triggered injection scheme. The plasma density profile (green solid) was measured using a Mach-Zehnder interferometer. (b) Illustration of the experimental setup.}
\end{figure}

Figure~\ref{fig1}(a) schematically depicts the injection and acceleration process. The tailored density profile features a plateau followed by a hump. As the laser pulse propagates through the density plateau, K-shell electrons (denoted $e_{ii}$) of nitrogen are released and trapped through ionization-induced injection \cite{pak2010injection,mcguffey2010ionization}. Once sufficient charge is loaded, further injection is suppressed due to beam loading \cite{tzoufras2008beam,tzoufras2009beam,couperus2017demonstration,zeng2014self}. 
Since the wakefield phase velocity is slightly lower than the velocity of the accelerated electron bunch after gaining sufficient energy \cite{lu2007generating}, the $e_{ii}$ bunch drifts forward relative to the wakefield until the laser pulse enters the density hump. Within the hump, the wakefield contracts, counteracting the electron dephasing \cite{guillaume2015electron,liu2024scalable}. The $e_{ii}$ electrons are re-accelerated to higher energy and emit brighter betatron X-rays in the shock-front plasma compared with a uniform-density plasma. At the falling edge of the density hump, the plasma wakefield expands, triggering the injection of additional electrons (denoted $e_{sf}$) from both hydrogen and nitrogen into the wakefield rear through shock-front injection. These $e_{sf}$ electrons further contribute to the X-ray emission.

\section{Experimental results}
The experiment was performed with a 40-TW-class Ti:Sapphire laser, as shown in Fig.~\ref{fig1}(b). Linearly polarized laser pulses with a central wavelength of 800 nm, a full width at half maximum (FWHM) duration of 37 fs, and energy of 1.1 J were focused to a 10-$\mu$m (FWHM) spot by an f/6 off-axis parabolic mirror. Approximately 51\% of the energy was enclosed within the Gaussian-fitted focal spot, corresponding to a peak intensity of $1.2\times10^{19}$ W/cm$^2$ and a normalized vector potential $a_0=2.4$ on target. The target was a 3-mm-diameter gas jet composed of 96\% hydrogen and 4\% nitrogen. To generate the desired structured density profile, an adjustable razor blade was mounted perpendicularly to the supersonic gas flow, forming a shock front with a density hump. The plasma electron density along the laser propagation axis was measured online using a Mach-Zehnder interferometer. A representative density profile is shown in Fig.~\ref{fig1}(a). 

Electron spectra were measured by a spectrometer consisting of a 1 T, 140-mm-long dipole magnet and a Gd$_2$S$_2$O:Tb phosphor screen (Lanex), with the visible fluorescence imaged onto a CMOS camera. Betatron X-rays were detected by a lead-shielded X-ray camera (ANDOR DO420-BN) placed 1.2 m downstream of the gas jet in vacuum. A 10-$\mu$m-thick aluminum foil in front of the detector blocked residual laser light. 
The X-ray spectrum was reconstructed using a Ross filter array \cite{guo2019enhancement} consisting of 5 $\mu$m Ti / 35 $\mu$m Al, 10 $\mu$m Cu / 40 $\mu$m Ti, and 20 $\mu$m Mo / 60 $\mu$m Cu pairs. The absolute photon yield was retrieved by accounting for the filter transmission, detector quantum efficiency, finite detector acceptance, and the fitted two-dimensional X-ray angular distribution. The complete reconstruction procedure and uncertainty analysis are provided in the Supplement 1, S1.

\begin{figure}[htb]
\centering\includegraphics{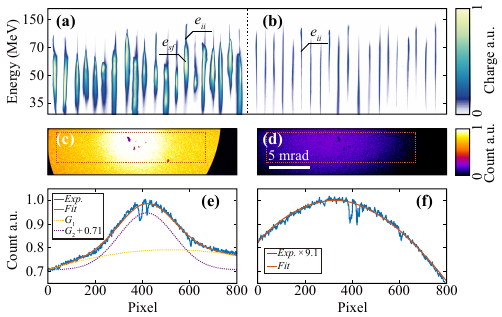}
\caption{\label{fig2} \textbf{Generation of dual-bunch electron and X-ray beams.} Left: shock-front-plasma configuration. Right: uniform-plasma configuration. 
(a,b) Twenty-shot electron spectra.
(c,d) Representative uncropped single-shot betatron X-ray profiles recorded over the full detector width.
(e,f) Corresponding horizontal lineouts (extracted from the regions marked by the orange dashed boxes) and their fits. For the shock-front-plasma configuration, the X-ray profile was fitted by a double-Gaussian function (orange solid) with RMS divergences of 25 mrad (yellow dashed, G$_1$) and 3 mrad (purple dashed, G$_2$). The uniform-plasma configuration was fitted by a single-Gaussian function (orange solid) with an RMS divergence of 11 mrad.}
\end{figure}

For each plasma density, the uniform-plasma source was first optimized by adjusting the laser-focus position relative to the nozzle. This procedure ensured that the enhancement obtained after introducing the density-tailored structure was evaluated relative to an optimized baseline rather than an unoptimized uniform-plasma source.

For the optimized shock-front-plasma configuration, the density maximum was located at around $z_s=1.9$~mm (the distance from the density peak of the shock front to the entrance of the gas jet) with a plateau electron density of $9\times10^{18}$ cm$^{-3}$ in the undisturbed region. The 20-shot electron spectra recorded with and without the razor blade are compared in Figs.~\ref{fig2}(a) and~\ref{fig2}(b). In the shock-front-plasma case, two electron populations are clearly resolved: a lower-charge, lower-divergence component associated with ionization-induced injection and a higher-charge, higher-divergence component associated with shock-front injection. Averaged over the 20 analyzed shots, the total electron charge above 30 MeV was $908\pm176$~pC. The ionization-injected bunch carried $94\pm19$~pC, whereas the shock-front-injected bunch carried $817\pm173$~pC.

Figures~\ref{fig2}(c) and~\ref{fig2}(d) show representative uncropped single-shot X-ray profiles. In the shock-front-plasma configuration, it is clear that the X-ray profile displays a dual-divergence characteristic with a smaller-divergence spot superimposed on a larger-divergence spot, matching the dual-bunch feature observed in the electron spectra. The overall X-ray profile was fitted by a double-Gaussian function with RMS divergences of 25 mrad and 3 mrad horizontally, as shown in Fig.~\ref{fig2}(e). As comparison, the uniform-plasma configuration exhibited only a single Gaussian-like spot with much darker intensity, and its RMS divergence was around 11 mrad (Fig.~\ref{fig2}(f)).  Combining these structural features of the X-ray profile with the electron spectra suggests that two electron bunches are accelerated and emit betatron radiation within the shock-front plasma.

\begin{figure}[htb]
\centering\includegraphics{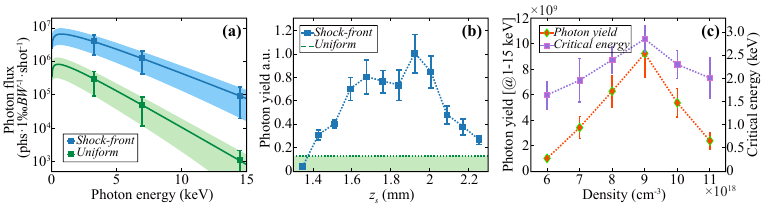}
\caption{\label{fig3} \textbf{Measured source performance and tunability.}
(a) Average betatron X-ray spectra for the shock-front-plasma and uniform-plasma configurations, including uncertainty bands.
(b) Photon yield versus shock-front position at a fixed plateau density of $9\times10^{18}$ cm$^{-3}$.
(c) Critical energy and integrated photon yield versus plasma density at $z_s=1.9$ mm. }
\end{figure}

Figure~\ref{fig3}(a) shows the X-ray spectra averaged over 20 shots. After correction for filter transmission, detector response, finite detector acceptance, and the fitted angular distribution, the shock-front-plasma configuration yields $(9.1\pm2.2)\times10^{9}$ photons/shot in the 1--15 keV range and $(3.1\pm0.9)\times10^{10}$ photons/shot over the full fitted spectrum. The corresponding enhancement relative to the optimized uniform-plasma source is approximately 8.5. Over the 0.5--7 keV range, the spectral photon flux exceeds $10^6$ photons/0.1\%BW/shot. 

The properties of the two X-ray pulses are coupled: both the emergence of the second pulse and the intensity enhancement are strictly dependent on the shock-front position and the plasma density. As shown in Fig.~\ref{fig3}(a), at a fixed plateau density of $9\times10^{18}$ cm$^{-3}$, the photon flux increases as the shock is moved into the principal emission region, remains close to its maximum for $1.6 \leq z_s \leq 2.0$ mm, and decreases when the shock is positioned too early or too late. Over this interval, the critical energy varies only moderately, from approximately 2.5 to 2.9 keV.  Furthermore, the longitudinal position of the shock front also determines the characteristics of the electron beam, most notably the charge and final energy, as detailed in the Supplement 1, S3.

At a fixed shock position of $z_s=1.9$ mm, increasing the plasma density from $6\times10^{18}$ to $1.1\times 10^{19}$ cm$^{-3}$ tunes the critical energy from 1.6 to 2.9 keV. The corresponding photon yield in the 1--15 keV range increases from $(9.7\pm2.1)\times10^{8}$ to $(9.1\pm2.2)\times10^{9}$ photons/shot, with an enhancement peaked at $n_e=9\times10^{18}$ cm$^{-3}$. Independent transverse-machining measurements confirm the principal longitudinal emission region and are presented in the Supplement 1, S2.

\section{Numerical simulations}
To identify the injection and radiation mechanisms, quasi-3D particle-in-cell (PIC) simulations were performed using OSIRIS \cite{fonseca2002osiris,li2021new} with longitudinal density profiles obtained from the experiment. The simulation windows had dimensions of $76.2\times50.8$ $\mu$m$^2$, with 12 azimuthal duplications and 3000 $\times$ 400 cells in the $z$ and $r$ directions, respectively. Each cell contained four macroparticles.

\begin{figure}[htb]
\centering\includegraphics{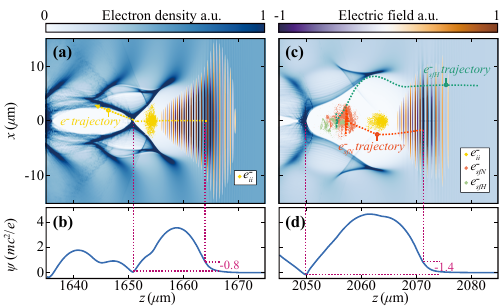}
\caption{\label{fig4} \textbf{Electron injection processes in the shock-front-plasma configuration.} Left: Simulation snapshots (a) and the corresponding pseudo-potential $\Psi$ (b) when the ionization-induced injection is truncated. Right: Simulation snapshots (c) and the corresponding $\Psi$ (d) as the laser pulse propagates through the falling edge of shock front structure.
}
\end{figure}

Figure~\ref{fig4}(a) shows the simulation results at $z\sim1500$ $\mu$m with the shock-front peak positioned at $z_s=1900$ $\mu$m. Here, $z$ represents the distance from the gas-jet entrance. The accelerating field near the rear of the bubble is reduced due to beam loading. As shown in Fig.~\ref{fig4}(b), the normalized pseudo-potential difference \cite{pak2010injection} ($\delta \Psi=\Psi_f-\Psi_b\approx-0.8>-1$) between the ionization point of the nitrogen inner-shell electrons ($\Psi_b$) and the wakefield rear ($\Psi_f$) does not reach the threshold required for ionization-induced injection. Ionized electrons from the inner shell of nitrogen fail to gain enough energy from the wakefield to be trapped, as shown by a representative electron trajectory in Fig.~\ref{fig4}(a). This condition suppresses further charge growth of the $e_{ii}$ electrons. This process occurs before the laser's interaction with the shock-front structure and is therefore common to both the uniform-plasma and shock-front-plasma cases.

At $z\sim2130$ $\mu$m, when the laser propagates through the descending gradient of the shock front (Fig.~\ref{fig4}(c)), another electron bunch ($e_{sf}$) is generated, with a temporal separation of $\sim$5 fs from the $e_{ii}$ bunch. The $e_{sf}$ electrons, containing background hydrogen electrons ($e_{sfH}$) and nitrogen inner-shell electrons ($e_{sfN}$), are produced by two schemes. The injection processes for both $e_{sfH}$ and $e_{sfN}$ electrons are illustrated by representative electron trajectories in Fig.~\ref{fig4}(c). With the expansion of the wakefield, the $e_{ii}$ bunch expels residual background electrons, thereby creating a larger and cleaner plasma bubble structure behind it. As shown in Fig.~\ref{fig4}(d), the $e_{sfN}$ electrons are injected as the condition for ionization-induced injection is met ($\delta \Psi\approx-1.4<-1$). During this process, the drive laser's $a_0$ increases by a factor of about 2 due to self-compression and self-focusing within the shock front \cite{nie2018relativistic}. This results in increased divergences and transverse sizes for $e_{sfN}$ electrons compared to the $e_{ii}$ electrons. Simultaneously, the $e_{sfH}$ electrons are trapped as the phase velocity of the wakefield decreased at the rear of the bubble. During injection, the $e_{sfH}$ electrons are rapidly decelerated transversely when arriving at the tail of the wakefield, which reduces their sizes and divergences \cite{xu2017high,hue2023control}.

\begin{figure}[htb]
\centering\includegraphics{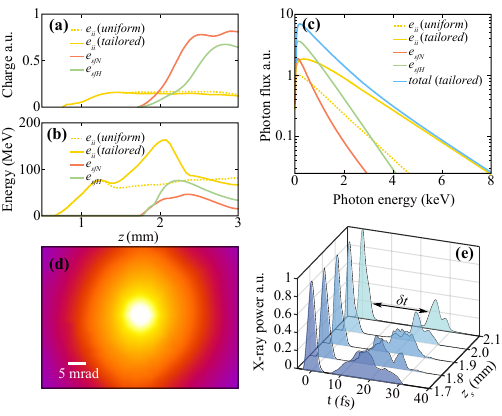}
\caption{\label{fig5} 
\textbf{Characterization of the simulated electrons and betatron X-rays.} (a) and (b) Evolution of electron charge and mean energy for shock-front plasma (solid) and uniform plasma (dashed). (c) Simulated X-ray spectra for shock-front plasma (solid) and uniform plasma (dashed). (d) Simulated X-ray profile for shock-front plasma. (e) Temporal distribution of simulated on-axis betatron X-ray power for different shock-front positions.
}
\end{figure}

To investigate the reason for the enhancement of betatron radiation in the shock-front-plasma configuration, the evolution of electron charge and mean energy is quantified in Figs.~\ref{fig5}(a) and~\ref{fig5}(b), and the corresponding betatron X-ray spectra calculated from the electron trajectories are shown in Fig.~\ref{fig5}(c). As the laser pulse enters the rising edge of the shock-front structure, the wakefield contracts, enabling the $e_{ii}$ electrons to experience stronger accelerating fields. Compared to the uniform-plasma case, the $e_{ii}$ electrons gain more energy over the same acceleration length, thereby emitting brighter and harder X-rays (a nearly 3.8-fold enhancement in photon yield above 1 keV). The charge of $e_{ii}$ electrons in the shock-front-plasma case is only slightly lower than that in the uniform-plasma case. However, positioning the shock front closer to the gas jet entrance leads to substantial electron loss due to inadequate drift distance before plasma bubble contraction, which in turn results in sharply reduced betatron radiation, as experimentally observed in Fig.~\ref{fig3}b.
After the density drops near the plasma exit, the energy of the $e_{ii}$ bunch finally decreases to the level similar to that observed in the uniform-plasma case, consistent with the experimental measurements shown in Figs.~\ref{fig2}(a) and~\ref{fig2}(b). For the $e_{sf}$ electrons, at the falling edge of the shock front, the $e_{sfH}$ and $e_{sfN}$ electrons are generated essentially simultaneously. Due to the injection positions of the $e_{sfH}$ electrons being closer to the rear of the wakefield, they gain higher energy than $e_{sfN}$ electrons, resulting in more intense X-ray radiation. However, compared to $e_{ii}$ electrons, the lower energies of both $e_{sfH}$ and $e_{sfN}$ electrons restrict emission to softer X-rays. Owing to their substantially higher charge ($e_{sfH}$ about 5 times the charge of $e_{ii}$, $e_{sfN}$ about 6 times the charge of $e_{ii}$), the total $e_{sf}$ electrons yield 2.6 times more X‑ray photons above 1 keV than the $e_{ii}$ electrons in the uniform-plasma case. Overall, the total X-ray yield in the shock-front-plasma configuration exceeds that of the uniform-plasma case by more than a factor of 6, closely matching experimental measurements.

Figure~\ref{fig5}(d) shows the simulated total far-field X-ray profile under the shock-front-plasma configuration, characterized by an intense, small spot superimposed on a larger spot. The larger spot with an RMS divergence of 19 mrad is emitted by the $e_{sf}$ electrons, while the smaller one with an RMS divergence of 2.5 mrad is generated by the $e_{ii}$ electrons. In contrast, the simulation of the uniform-plasma case shows an $e_{ii}$-generated X-ray divergence of 9.5 mrad (RMS). These simulation results are in good agreement with the experimentally measured results shown in Figs.~\ref{fig2}(c)--(f).

The temporal separation $\delta t$ of the two radiation components was not measured directly in the present experiment and is instead inferred from the PIC simulations. The simulations show that the shock-front position controls the relative injection phase of the two electron bunches and hence their radiation delay. 
As the wakefield propagates at a phase velocity $v_{\phi}\simeq c[1-3\omega_p^2/(2\omega_0^2)]$ ($\omega_p$ represents the plasma frequency, and $\omega_0$ represents the laser frequency) \cite{lu2007generating}, the $e_{ii}$ bunch drifts forward relative to the wake by $\delta l=l[(c-v_{\phi})/v_{\phi}]$ over a propagation distance $l$. For a plasma density of $9\times10^{18}$ cm$^{-3}$, an additional 100 $\mu$m of wake propagation (equivalently, a 100 $\mu$m increase in $z_s$) results in a drift of $\delta l\approx$ 0.8 $\mu$m, corresponding to an increase in $\delta t$ of 2.7 fs per 100 $\mu$m. Figure~\ref{fig5}(e) shows the simulated distribution of on-axis ($\theta$ = 0) X-ray power for $z_s$ tuning from 1.7 to 2.1 mm. The $e_{ii}$ X-ray pulses exhibit shorter durations ($\sim5$ fs FWHM) and higher peak intensity compared to the $e_{sf}$ pulses, with a maximum $\delta t$ of $\sim20$ fs. The simulations indicate that for every 100 $\mu$m increase in $z_s$, $\delta t$ increases by $\sim3.1$ fs on average, which is in close agreement with the estimation. 

Simulations incorporating representative fluctuations of $\pm1\%$ in laser energy and $\pm5\%$ in plasma density indicate that the temporal separation between the two pulses is insensitive to the investigated laser-energy variation, whereas the density variation produces an approximately $\pm10\%$ change in the pulse separation (see the Supplement 1, S5 for details).

We also note that direct measurement of the femtosecond X-ray temporal structure remains challenging. A possible future approach is femtosecond relativistic electron microscopy (FREM) \cite{wan2022direct,wan2023femtosecond,wan2024real,zhang2017femtosecond}, in which a femtosecond electron probe images the in situ structure of the accelerated electron beams within the plasma bubble. The measured bunch separation could then be used, together with radiation modeling, to constrain the relative timing of the two X-ray-emission components. A proof-of-principle simulation of this diagnostic is provided in the Supplement 1, S4.

\section{Discussion}
For downstream applications of such dual X-ray pulses, a large-aperture grazing-incidence X-ray mirror could be used to collect and refocus part of the emitted radiation, while an external X-ray optical delay line could extend the accessible delay without introducing the electron-beam dispersion associated with a magnetic chicane. Based on the estimated collection and transport efficiency, the total on-target photon number could reach the $10^9$-photons/shot level. This estimate suggests potential compatibility with broadband X-ray absorption spectroscopy (XAS), although the usable photon number in each pulse will depend on the focusing geometry, spectral selection, and spatial overlap. The temporal order of the two components is fixed, with the first pulse exhibiting a harder spectrum and smaller divergence, making the source more naturally suited to broadband XAS than to experiments requiring two independently tunable monochromatic pulses. Two possible experimental configurations, including a direct short-delay pump-probe arrangement and an extended-delay configuration incorporating an X-ray optical delay line, are presented and discussed quantitatively in the Supplement 1, S6.

In conclusion, we have experimentally demonstrated high-flux betatron radiation associated with two electron populations generated in a density-tailored gas-mixture target driven by a 40-TW-class laser. The measured electron spectra and dual-component X-ray angular profiles, supported by PIC simulations, show that ionization-induced and shock-front-triggered injections generate two distinct electron populations that contribute to the radiation. Rephasing acceleration of the first bunch, together with the high charge of the second bunch, increases the photon yield to $\sim10^{10}$ photons/shot. The simulations predict few-femtosecond X-ray pulse durations and a separation tunable up to approximately 20 fs through adjustment of the shock-front position. Although direct temporal characterization and an application experiment remain to be demonstrated, the measured source performance provides a basis for compact dual-pulse ultrafast X-ray spectroscopy.

\begin{backmatter}
\bmsection{Funding}
Strategic Priority Research Program of the Chinese Academy of Sciences (XDB0530000);
IHEP Talent Introduction Program (E65152U1);
National Natural Science Foundation of China (12574380 and 12405169);
Key Scientific Research Projects of Henan Provincial Colleges and Universities (25ZX002);
Natural Science Foundation of Henan Province (252300421300);
Science Fund Program for Distinguished Young Scholars of the National Natural Science Foundation of China (Overseas);
Discipline Construction Foundation of ``Double World-class Project''.

\bmsection{Acknowledgment}
The simulation work was supported by Center of High Performance Computing, Tsinghua University. 

\bmsection{Disclosures}
The authors declare no conflicts of interest.

\bmsection{Data availability} Data underlying the results presented in this paper are not publicly available at this time but may be obtained from the authors upon reasonable request.

\bmsection{Supplemental document}
See Supplement 1 for supporting content.
\end{backmatter}

\bibliography{refs}

\newpage
\appendix

\title{Bright dual-pulse betatron X-ray generation from a laser wakefield accelerator: supplemental document}
\setcounter{figure}{0}
\renewcommand{\thefigure}{S\arabic{figure}}
\captionsetup[figure]{labelfont=bf}
\section{X-ray spectrum reconstruction}

\begin{figure}[h]
\includegraphics[width=1\linewidth]{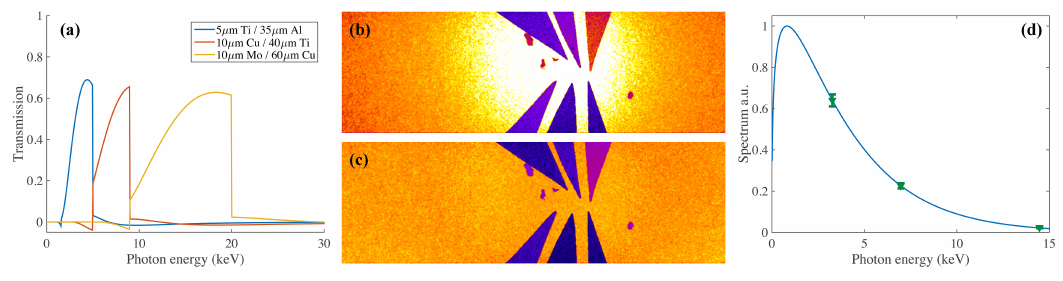}
\caption{\label{figS1} Reconstruction of the betatron X-ray spectrum. (a) Transmission curves for the Ross filter pairs. (b) Typical attenuated X-ray image recorded through the filters. (c) Image after flat-field correction. (d) Fitted on-source single-shot betatron X-ray spectrum.}
\end{figure}

Figure~\ref{figS1}(a) shows the transmission curves of the Ross filter pairs, and Figure~\ref{figS1}(b) displays a typical attenuated image through the filters. By applying a dual-Gaussian fit to extract the X-ray profile for flat-field correction (Fig.~\ref{figS1}(c)), we can obtain the transmission through individual filters. Accounting for the transmission of the 10 $\mu$m aluminum foil and the quantum efficiency of the X-ray camera, the on-source spectrum can be fitted using a synchrotron spectrum (Fig.~\ref{figS1}(d)). 

The uncertainty in the critical energy primarily originates from statistical fluctuations in the photon counts within each filter region, yielding a fitted critical energy of $2.86\pm0.06$ keV. Furthermore, extracting the $\pm1\sigma$ confidence intervals of the Gaussian parameters from the X-ray profile fit results in a relative error of approximately 11\% for the total counts due to the finite acceptance. Combining the spectrum deposited on the camera and the fitted X-ray profile, we determine the photon number for this shot to be $(2.9\pm0.3)\times10^{10}$ photons/shot integrated over the entire spectral range. 

To ensure that the photon flux did not exceed the saturation limit of the X-ray camera, we estimated the corresponding saturation threshold. Multiplying the source spectrum by the transmission of the 10 $\mu$m Al foil and the quantum efficiency curve of the X-ray camera, we calculate that each photon emitted from the source creates, on average, 43 electron-hole pairs. To reach the full well capacity (500,000 e$^-$) of a single pixel, that specific pixel must receive approximately $1.2\times10^4$ photons. In our experimental setup, the acceptance angle of the central pixel is 0.022 mrad. Based on this, we calculate that the single-shot on-source photon yield required to saturate the central pixel is approximately $8\times10^{10}$ photons, which is safely above our experimentally measured single-shot photon yield.

\section{Diagnostics of X-ray emission}
\begin{figure}[h]
\includegraphics[width=1\linewidth]{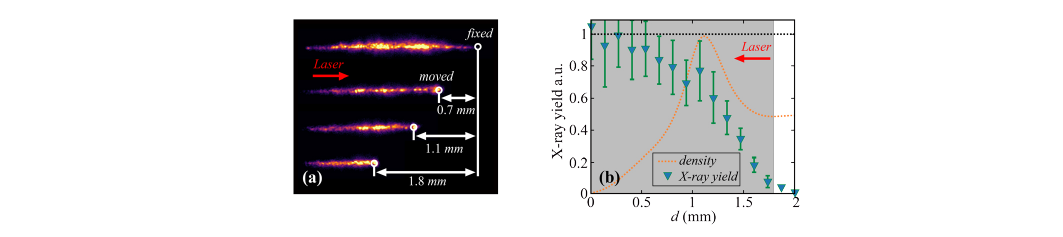}
\caption{\label{figS2} Diagnostics of the betatron radiation region. (a) Thomson side-scattering images of the plasma ablated using the transverse laser machining method. By shifting the starting position while keeping the endpoint fixed, the modified region was scanned. (b) Measured X-ray yield against the machined length for the uniform-plasma configuration. Each data point is an average of 10 shots. The shaded region indicates the emission zone of $e_{ii}$ betatron X-rays. The black dashed line represents the X-ray yield without transverse machining, and the orange dashed line shows the shock-front density profile for $z_s=1.9$ mm.}
\end{figure}

We characterized the betatron radiation emission process in the uniform-plasma case using the transverse laser machining method. A tunable, perpendicularly propagating line-focus laser beam was used to selectively ablate the plasma formed in the gas target, thereby disrupting the radiation emission from the modified regions. By scanning the length $d$ of the modified region, as verified by Thomson side-scattering diagnostics (Fig.~\ref{figS2}(a)), the dynamics of the X-ray emission can be mapped. Figure~\ref{figS2}(b) plots the dependence of the X-ray yield on $d$ for the uniform-plasma case, alongside a representative tailored plasma density profile at $z_s=1.9$ mm. The $e_{ii}$ X-rays were found to be predominantly emitted within the region of $d<1.7$ mm. Consequently, to effectively utilize the shock front without truncating this primary emission, it should be positioned at $z_s>1.3$ mm, which is consistent with the experimental observations.

\section{Electron spectrum behavior}
\begin{figure}[htb]
\includegraphics[width=1\linewidth]{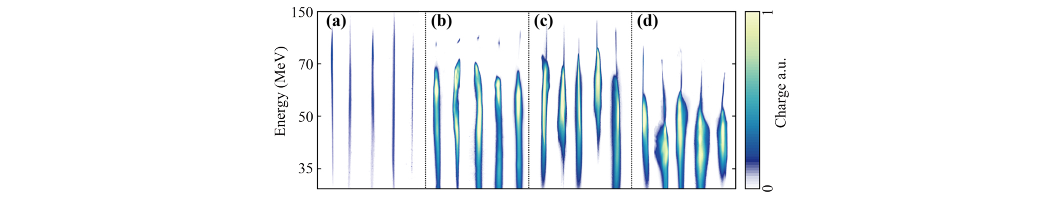}
\caption{\label{figS3} Electron spectra versus the shock-front position. The spectra were measured with a fixed plasma density of $9\times10^{18}$ cm$^{-3}$ for the uniform-plasma case (a), and for shock-front-plasma case with $z_s$ = 1.5 mm (b), 1.9 mm (c), and 2.1 mm (d).}
\end{figure}

In the shock-front-plasma configuration, the electron beam properties, particularly the charge and spectral structure, depend on the position of the shock front $z_s$. Figure~\ref{figS3} compares typical electron spectra measured for the uniform-plasma case and for three different values of $z_s$ in the shock-front-plasma case, where the plateau plasma density $n_e$ of the undisturbed region was $9\times10^{18}$ cm$^{-3}$.  

We observe that when the shock front is introduced at an early position (smaller $z_s$), the charge of the $e_{ii}$ bunch drops significantly, but its monoenergetic quality improves. Conversely, when the shock front is introduced at a later position (larger $z_s$), the total charge shows no significant variation, but the exit energies of both the $e_{ii}$ and $e_{sf}$ bunches decrease.

\section{Diagnostics of the pulse temporal structure using FREM}
\begin{figure}[h]
\includegraphics[width=1\linewidth]{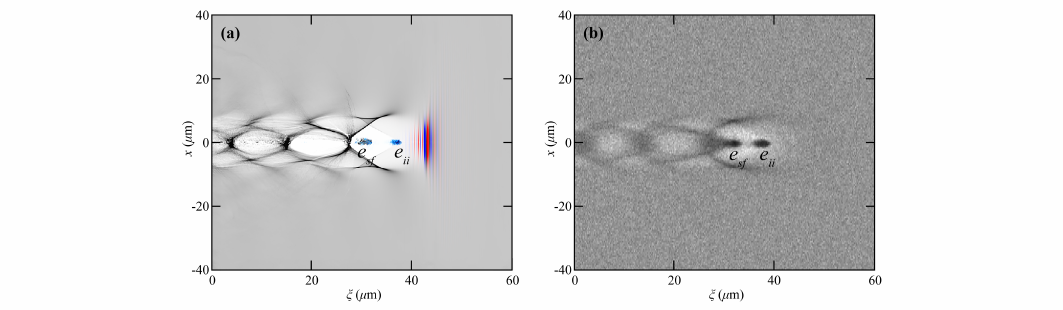}
\caption{\label{figS4} FREM simulation. (a) Simulation snapshot of the density-tailored case. (b) Simulated probe relative density. The probe electron bunch features an energy of 400 MeV and an ultrashort pulse duration of 2 fs.}
\end{figure}

A femtosecond relativistic electron microscopy (FREM) method was proposed to characterize the temporal structure of the dual‑pulse X‑ray source. This approach utilizes a femtosecond electron probe to capture, \textit{in situ}, the accelerated electron‑beam structures inside the plasma bubble, from which the X‑ray temporal profile is indirectly inferred. Figure~\ref{figS4} presents a PIC simulation test of this approach, wherein a 400 MeV electron probe bunch with a duration of 2 fs crosses the wakefield and gets modulated. The resulting density distribution (Fig.~\ref{figS4}(b)) clearly reveals the dual accelerated electron bunches inside the bubble. 

\section{X-ray pulse delay jitter}
\begin{figure}[h]
\includegraphics[width=1\linewidth]{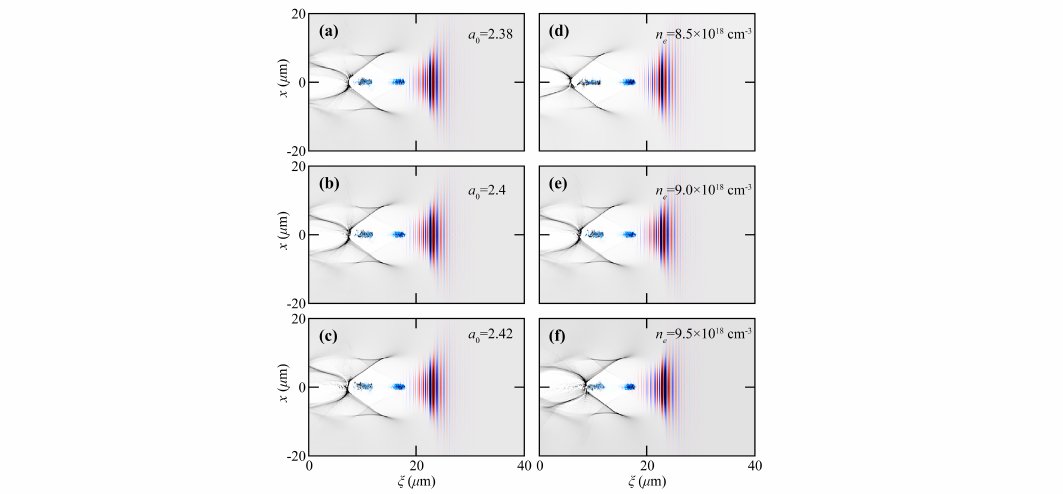}
\caption{\label{figS5} PIC simulations assessing the impact of typical experimental fluctuations on the inter-pulse delay. Variations were introduced to the driving laser energy (panels (a)–(c)) and the plasma density (panels (d)–(f)). The results reveal that given current experimental capabilities, the temporal separation is highly robust against laser energy jitter, whereas plasma density fluctuations induce a manageable variation of $\sim\pm10\%$ (9.9--12.1 fs).}
\end{figure}

The stability of the pulse delay has been evaluated using PIC simulations. We introduced typical jitter into the driving laser energy ($\sim\pm1\%$, Figs.~\ref{figS5}(a)--(c)) and the background plasma density ($\sim\pm5\%$, Figs.~\ref{figS5}(d)--(f)) to mimic the fluctuation of experimental conditions. The simulations show that, under currently achievable experimental stability, plasma density jitter exerts a relatively more pronounced impact on the temporal separation, inducing a variation of $\sim\pm$10\% (ranging from 9.9 to 12.1 fs). In contrast, the influence of laser energy fluctuations is negligible.

\section{Potential pump-probe measurement configuration}
\begin{figure}[h]
\includegraphics[width=1\linewidth]{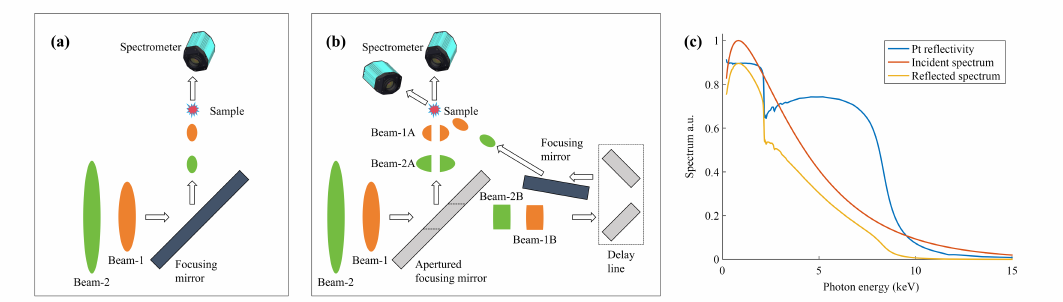}
\caption{\label{figS6} Experimental configurations for dual-pulse ultrafast X-ray absorption spectroscopy. (a) A standard pump-probe setup utilizing a single focusing mirror. Beam-1 acts as the pump to trigger the ultrafast process, while Beam-2 serves as the probe. The pump-probe delay can be tuned over tens of femtoseconds by adjusting the shock-front position, with the transient spectrum reconstructed via steady-state baseline subtraction. (b) An extended configuration enabling wide-range delay tuning and `one-pump, two-probes' measurements. An apertured X-ray optic spatially divides the incident beam into Beam-A and Beam-B. Beam-A probes short-delay dynamics similarly to setup (a). Meanwhile, an X-ray optical delay line adjusts the relative timing of Beam-B. Because Beam-1B possesses a tighter divergence and harder spectrum, the pulse is effectively `purified' at high photon energies ($>2$ keV), allowing for the extraction of dynamic information at significantly larger temporal delays. (c) Calculated reflectivity of the Pt-coated mirror at a grazing incidence angle of 10 mrad. For a typical betatron X-ray spectrum, the effective reflectivity of this mirror is approximately 84\%.}
\end{figure}

For X-ray pump-probe experiments, we propose two possible configurations for ultrafast X-ray absorption spectroscopy, as shown in Fig.~\ref{figS6}.

Configuration (a) (see Fig.~\ref{figS6}(a)): Both Beam-1 and Beam-2 are focused onto the sample by a single mirror. Beam-1 operates as the pump to trigger the ultrafast process, and Beam-2 serves as the probe to capture the evolving dynamics. By subtracting the steady-state baseline (the signal inherent to Beam-1), the transient spectrum captured by Beam-2 can be reconstructed. The pump-probe delay can be tuned over tens of femtoseconds by shifting the shock-front position. 

Configuration (b) (see Fig.~\ref{figS6}(b)): This setup enables a wider delay tuning range and a `one-pump, two-probes' measurement. An X-ray optic with a central aperture splits the incident beam into two paths: Beam-A (containing Beam-1A and Beam-2A) and Beam-B (containing Beam-1B and Beam-2B). An X-ray optical delay line adjusts the relative timing between Beam-A and Beam-B over a wide range without significantly broadening the pulse duration. Beam-A probes short-delay dynamics as in (a). For Beam-B, because Beam-1B has a tighter divergence and harder spectrum, its photon flux at high energies ($>2$ keV) exceeds that of Beam-2B by more than two orders of magnitude, effectively `purifying' the pulse in that energy range. This allows extraction of dynamic information at much larger delays.

To estimate the photon number delivered to the sample, we modeled the use of a large-aperture platinum (Pt)-coated ellipsoidal or toroidal mirror (e.g., $1000\times100$ mm$^2$) operating at a grazing incidence angle of $\sim10$ mrad. For a typical betatron broadband spectrum, the effective reflectivity of this mirror reaches $\sim84\%$ (Fig.~\ref{figS6}(c)). Placed 500 mm downstream from the source, this geometric configuration provides an effective collection angle of $20\times200$ mrad$^2$. Given the $\sim25$ mrad divergence of the betatron radiation, the overall transport efficiency, accounting for both geometric collection and mirror reflectivity, is approximately 25\%. Assuming an initial photon flux on the order of 10$^{10}$ photons per shot, an effective flux of $\sim10^9$ photons per shot can be delivered to the sample following optical transport.

Furthermore, the flux and spectra of the two X-ray components are coupled. Both the enhancement of the first pulse and the generation of the second pulse depend on the shock-front position and plasma density. 
This coupling does not necessarily hinder the pump-probe experiments. The source is broadband, and the photon flux remains high over a finite range of shock positions and plasma densities. For X-ray absorption spectroscopy, one can select an appropriate spectral window and normalize the measured transient spectrum by subtracting the steady-state baseline. 

\end{document}